\documentclass[aps,pra,reprint,superscriptaddress]{revtex4-2}

\usepackage{graphicx}
\usepackage{tabularx}
\usepackage{color}

\usepackage{xspace}

\begin{document}

\title{Irradiation driven molecular dynamics simulation of the FEBID process for Pt(PF$_3$)$_4$}

\author{Alexey Prosvetov}
\email{prosvetov@mbnexplorer.com}
\affiliation{MBN Research Center, Altenh\"oferallee 3, 60438 Frankfurt am Main, Germany}
\author{Alexey V. Verkhovtsev}
\email{verkhovtsev@mbnexplorer.com}
\affiliation{MBN Research Center, Altenh\"oferallee 3, 60438 Frankfurt am Main, Germany}
\affiliation{On leave from Ioffe Physical-Technical Institute, Polytekhnicheskaya 26, 194021 St. Petersburg, Russia}
\author{Gennady Sushko}
\affiliation{MBN Research Center, Altenh\"oferallee 3, 60438 Frankfurt am Main, Germany}
\author{Andrey V. Solov'yov}
\affiliation{MBN Research Center, Altenh\"oferallee 3, 60438 Frankfurt am Main, Germany}
\affiliation{On leave from the Ioffe Physical-Technical Institute, Polytekhnicheskaya 26, 194021 St. Petersburg, Russia}

\begin{abstract}
This paper presents a detailed computational protocol for atomistic simulation of the formation and growth of metal-containing nanostructures during the Focused Electron Beam Induced Deposition (FEBID) process. The protocol is based upon the Irradiation-Driven Molecular Dynamics (IDMD) -- a novel and general methodology for computer simulations of irradiation-driven transformations of complex molecular systems by means of the advanced software packages MBN Explorer and MBN Studio. Atomistic simulations performed following the formulated protocol provide valuable insights into the fundamental mechanisms of electron-induced precursor fragmentation and the related mechanism of nanostructure formation and growth using FEBID, which are essential for the further advancement of FEBID-based nanofabrication. The developed computational methodology is general and applicable to different precursor molecules, substrate types, irradiation regimes, etc. The methodology can also be adjusted to simulate the nanostructure formation by other nanofabrication techniques using electron beams, such as direct electron beam lithography. In the present study, the methodology is applied to the IDMD simulation of the FEBID of Pt(PF$_3$)$_4$ -- a  widely studied  precursor  molecule -- on a SiO$_2$ surface. The simulations reveal the processes driving the initial phase of nanostructure formation during FEBID, including nucleation of Pt atoms, formation of small metal clusters on the surface, followed by their aggregation and the formation of dendritic platinum nanostructures. The analysis of the simulation results provides space resolved relative metal content, height and the growth rate of the deposits which represent a valuable reference data for the experimental characterization of the nanostructures grown by FEBID.
\end{abstract}


\maketitle

\section{Introduction}
\label{Intro}

The controllable fabrication of nanostructures with nanoscale resolution remains a considerable scientific and technological challenge \cite{Cui_Nanofabrication_book}. To address such a challenge, novel techniques exploiting irradiation of nanosystems with collimated electron and ion beams have been developed \cite{Utke_book_2012, DeTeresa-book2020}.
One of such techniques is electron beam lithography (EBL), which is similar to conventional optical lithography but relies on the change of solubility after electron exposure of irradiation-sensitive resists. The EBL process includes the surface coating of a resist, exposure to the energetic electron beam, and further development of the surface to remove irradiated or non-irradiated material.
Another technique, Focused Electron Beam Induced Deposition (FEBID) \cite{VanDorp2008, Utke_book_2012,  DeTeresa-book2020, Huth2012}, is based on the irradiation of precursor molecules \cite{Barth2020_JMaterChemC} by high-energy electrons while they are being adsorbed upon a substrate.
Electron-induced decomposition releases the non-volatile part of the precursor molecules, forming a deposit on the surface, whereas the volatile fragments are pumped out of the working chamber. Nowadays, FEBID permits the fabrication of nanostructures with the size down to a few nanometers, which is similar to the size of the incident electron beam \cite{Utke2008}.

To date, FEBID has mainly relied on precursor molecules developed for chemical vapor deposition (CVD) -- a process mainly governed by thermal decomposition, while dissociation mechanisms in FEBID are predominantly electron-induced reactions.
While primary electron (PE) energies during FEBID are typically between 1 keV and 30 keV, chemical dissociation is most efficient for low-energy (up to several tens of eV) secondary electrons (SE) created in large numbers when a high-energy PE beam impinges on a substrate. Secondary electrons are emitted from the substrate and the deposit, making the electron-induced chemistry that governs FEBID substantially complicated.
As a result, the precise control of the size, shape and chemical composition of the fabricated nanostructures is still a technological challenge \cite{Utke2008}, mainly originating from the lack of molecular-level understanding of irradiation-driven chemistry (IDC) underlying nanostructure formation and growth.

Further advances in FEBID-based nanofabrication require a deeper understanding of the relationship between deposition parameters and physical characteristics of fabricated nanostructures (size, shape, purity, crystallinity, etc.).
Further advancement of the existing experimental techniques and molecular-level computational modeling can provide insights into the fundamental mechanisms of electron-induced precursor fragmentation and the corresponding mechanism of nanostructure formation and growth using FEBID.

Until recently, most computer simulations of FEBID and the nanostructure growth have been performed using a Monte Carlo approach and diffusion-reaction theory \cite{Utke_book_2012, Fowlkes2010, Sanz-Hernandez2017, Toth2015}, which allow simulations of the average characteristics of the process concerning local growth rates and the nanostructure composition. However, these approaches do not provide any molecular-level details regarding structure (crystalline, amorphous, mixed) and the IDC involved.
At the atomic level, quantum chemistry methods have been utilized to analyze the adsorption energies and optimized structures of different precursor molecules deposited on surfaces \cite{Muthukumar2012, Muthukumar2018}. Nevertheless, ab initio methods are applicable to relatively small molecular systems with a typical size of up to a few hundred atoms. This makes ab initio approaches of limited use to describe the irradiation-induced chemical transformations occurring during the FEBID process.

A breakthrough into the atomistic description of FEBID has been achieved recently by means of Irradiation-Driven Molecular Dynamics (IDMD) \cite{Sushko2016}, a novel and general methodology for computer simulations of irradiation-driven transformations of complex molecular systems. This approach overcomes the limitations of previously used computational methods and describes FEBID-based nanostructures at the atomistic level by accounting for quantum and chemical transformation of surface-adsorbed molecular systems under focused electron beam irradiation \cite{Sushko2016, MBNbook_Springer_2017, DeVera2020, EPJD_IDMD-review2021}.

Within the IDMD framework various quantum processes occurring in an irradiated system (e.g. ionization, bond dissociation via electron attachment, or charge transfer) are treated as random, fast and local transformations incorporated into the classical MD framework in a stochastic manner with the probabilities elaborated on the basis of quantum mechanics \cite{Sushko2016}.
Major transformations of irradiated molecular systems
(e.g. molecular topology changes, redistribution of atomic partial charges, alteration of interatomic interactions, etc.)
are simulated by means of MD with reactive force fields \cite{Sushko2016a, Friis2020} using the advanced software packages MBN Explorer \cite{Solovyov2012} and MBN Studio \cite{Sushko2019}. MBN Explorer is a multi-purpose software package for multiscale simulations of structure and dynamics of complex Meso-Bio-Nano systems \cite{MBNbook_Springer_2017}. MBN Studio is a powerful multi-task toolkit enabling to set up and start MBN Explorer calculations, to monitor their progress, to examine calculation results, to visualize inputs and outputs, and to analyze specific characteristics determined by the output of MD simulations \cite{Sushko2019}.

In the pioneering study \cite{Sushko2016} IDMD was successfully applied for the simulation of FEBID of W(CO)$_6$ precursors on a SiO$_2$ surface and enabled to predict the morphology, molecular composition and growth rate of tungsten-based nanostructures emerging on the surface during the FEBID process.
The follow-up study \cite{DeVera2020} introduced a novel multiscale computational methodology that couples Monte Carlo simulations for radiation transport with IDMD for simulating the IDC processes with atomistic resolution. The spatial and energy distributions of secondary and backscattered electrons emitted from a SiO$_2$ substrate were used to simulate electron-induced formation and growth of metal nanostructures after deposition of W(CO)$_6$ precursors on SiO$_2$.


Investigation of the physicochemical phenomena that govern the formation and growth of nanostructures coupled to radiation is a complex multi-parameter problem. Indeed, different precursor molecules, substrate types, irradiation, replenishment and post-processing regimes, and additional molecular species that may facilitate precursor decomposition can be explored to improve the purity of grown deposits and increase the deposition rate. Therefore, it is essential to develop a comprehensive computational protocol for atomistic simulations of electron-induced nanostructure formation during the FEBID process.

This paper outlines a detailed computational methodology for modeling the formation and growth of metal-containing nanostructures during FEBID by means of IDMD, which was developed in the previous studies \cite{Sushko2016, DeVera2020}.  Different computational aspects of the methodology and the key input parameters describing the precursor molecules, the substrate, and the irradiation and replenishment conditions are systematically described. A step-by-step simulation and parameter determination workflow represents a comprehensive computational protocol for simulating and characterizing a broad range of nanostructures created by means of FEBID. The presented methodology has been developed with a focus on the FEBID process using a pulsed electron beam. However, it can be adjusted and extended by including processes relevant to different EBL or FEBID regimes, e.g. variation of the replenishment rates, concentration and distribution of molecules added and removed from the system, etc.

The formulated computation protocol is applied to simulation of the FEBID of Pt(PF$_3$)$_4$ -- a widely studied precursor molecule \cite{Wang2004a, Barry2006a, Botman2009, Landheer2011, May2012a, Zlatar2016} -- on a fully hydroxylated SiO$_2$ surface.
As such, this work extends the earlier IDMD-based studies \cite{Sushko2016, DeVera2020} of the FEBID of W(CO)$_6$ towards another precursor molecule that has been commonly used to fabricate platinum-containing nanostructures.
In contrast to the earlier studies \cite{Sushko2016, DeVera2020} we consider the case of low precursor surface coverage (below one monolayer), in which surface diffusion plays an important role in the formation of deposits.
Particular focus is made on the atomistic characterization of the initial stage of the FEBID process, including nucleation of Pt atoms, formation of small metal clusters on the surface followed by their aggregation and, eventually, the formation of dendritic platinum nanostructures.

\begin{figure*}[t!]
\centering
\includegraphics[width=0.65\textwidth]{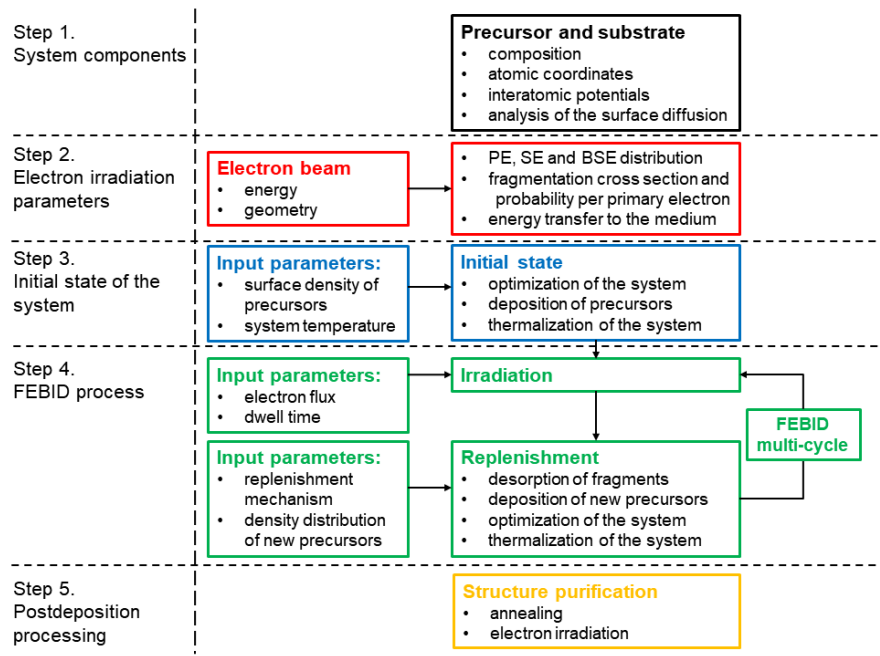}
\caption{Atomistic IDMD-based simulation protocol of the FEBID process by means of MBN Explorer \cite{Solovyov2012} and MBN Studio \cite{Sushko2019} software packages.}
\label{Fig:FEBID_protocol}
\end{figure*}

Morphology of the nanostructures grown during the FEBID process has not yet been thoroughly investigated on the atomistic level although it governs many physical properties such as electrical and thermal conductivity, and magnetic properties \cite{Huth_2009, Huth2020}. Atomistic simulations provide insights into the internal structure of the deposits and its evolution depending on the regimes of the FEBID process. In this study morphology, size and metal content of the deposited Pt-containing material are quantitatively analyzed as functions of electron fluence and adsorbate concentration.

\section{Computational methodology}
\label{Methods}

Computer simulations of the FEBID process of Pt(PF$_3$)$_4$ have been performed by means of the MBN Explorer software package \cite{Solovyov2012}. The MBN Studio toolkit \cite{Sushko2019} has been utilized to create the systems, prepare all necessary input files and analyze simulation outputs.

FEBID operates through successive cycles of precursor molecules replenishment on a substrate and irradiation by a tightly-focused pulsed electron beam, which induces the release of metal-free ligands and the growth of metal-enriched deposits. It involves a complex interplay of phenomena taking place on different temporal and spatial scales: (i) deposition, diffusion and desorption of precursor molecules on the substrate; (ii) transport of the primary, secondary and backscattered electrons; (iii) electron-induced dissociation of the adsorbed molecules; and (iv) the follow-up chemistry. Each of these phenomena requires dedicated computational and theoretical approaches.

The general workflow of the atomistic IDMD-based simulation of the FEBID process and the corresponding input parameters are summarized in  Figure~\ref{Fig:FEBID_protocol}.
The formulated procedure for the atomistic simulation of FEBID is general and applicable to any combination of the precursor, substrate and electron beam with minimum variations of the case-specific parameters.
The methodology can also be adjusted to simulate the nanostructure formation by other nanofabrication techniques using electron beams, such as direct-write EBL \cite{DeTeresa-book2020}.

In the following section, all steps of the protocol shown in Figure~\ref{Fig:FEBID_protocol} are described in detail, going from the system specification (Step~1) to the multi-cycle simulation of the FEBID process (Step~4) and simulation of post-deposition processing (Step~5).
At Step 1 one needs to choose the type of precursor molecules and the substrate, specify atomic coordinates and the parameters of interatomic potentials. At Step 2 the spatial and energy distributions of the electron irradiation field, the fragmentation cross section, as well as the energy deposited into the system during the fragmentation process are specified. The initial state of the adsorbed molecules to be exposed to electron-beam irradiation is created at Step~3. This follows by the multiple cycling of irradiation and replenishment phases at Step~4 using the experiment-related input parameters. Additional post-growth processing that enables purification of the deposited structures is simulated at Step~5. All these steps are described in greater details below.

\subsection*{Step 1. System components: precursor and substrate}
\label{Method:System_components}

The first step of the FEBID simulation procedure shown in Figure~\ref{Fig:FEBID_protocol} concerns the specification of a precursor molecule and a substrate. The selection of the system components is usually linked to available experiments.
Common experimentally used substrate materials are SiO$_2$, Si, Au, amorphous carbon, etc. A choice of the substrate affects the adsorption, desorption and diffusion of precursor molecules on the surface as well as the yields of secondary and backscattered electrons. These quantities affect the fragmentation rate of the adsorbed precursor molecules and hence the nanostructure growth rate.

The chemical composition and geometry of both the precursor and the substrate are specified using the standard .pdb or .xyz file formats. Atomic coordinates for many different precursor molecules can be found in the well-established online databases, e.g. NIST Chemistry WebBook (https://webbook.nist.gov/) and PubChem (https://pubchem.ncbi.nlm.nih.gov/), or determined via DFT calculations \cite{DeVera2019,Muthukumar2012,Muthukumar2018}.

MBN Explorer and MBN Studio enable the creation of various crystalline substrates for which the unit cell and translation vectors are specified. The software tools enable also creating amorphous substrates, e.g. amorphous silica or amorphous carbon, which are commonly used in FEBID and surface science experiments.

The structure of precursor molecules, their interaction with a substrate and the dynamics of nanostructure formation and growth are influenced by interatomic interactions between the system's constituents.

\subsubsection*{Interatomic potentials}

The precursor molecules are described via the reactive CHARMM (rCHARMM) force field \cite{Sushko2016a, Friis2020}. rCHARMM permits simulations of systems with dynamically changing molecular topologies, which is essential for modeling the precursor fragmentation and the formation of metal-containing nanostructures. A detailed description of rCHARMM is given in Ref.~\cite{Sushko2016a}, while its key aspects are summarized below.

The covalent bond interactions are described in rCHARMM by means of the Morse potential:
\begin{equation}
    U^{{\rm bond}}(r_{ij}) = D_{ij} \left[ e^{-2\beta_{ij}(r_{ij} - r_0)} - 2e^{-\beta_{ij}(r_{ij} - r_0)} \right] \ .
    \label{Eq. Morse}
\end{equation}
Here $D_{ij}$ is the dissociation energy of the bond between atoms $i$ and $j$, $r_0$ is the equilibrium bond length, and the parameter $\beta_{ij} = \sqrt{k_{ij}^{r} / D_{ij}}$ (where $k_{ij}^{r}$ is the bond force constant) determines the steepness of the potential. The bonded interactions are truncated at a user-defined cutoff distance that characterizes the distance beyond which the bond is considered broken and the molecular topology of the system changes. The bond energy given by Eq.~(\ref{Eq. Morse}) asymptotically approaches zero at large interatomic distances.

The rupture of covalent bonds in the course of simulation employs the following reactive potential for the valence angles \cite{Sushko2016a}
\begin{equation}
U^{{\rm angle}}(\theta_{ijk}) =
2 k^\theta_{ijk} \, \sigma(r_{ij}) \, \sigma(r_{jk}) \left[ 1 - \cos(\theta_{ijk}-\theta_0 )  \right] \ ,
\label{Eq. Angles}
\end{equation}
where $\theta_0$ is the equilibrium angle, $k^{\theta}$ is the angle force constant, and the sigmoid function
\begin{equation}
\sigma(r_{ij}) = \frac{1}{2} \left[1-\tanh(\beta_{ij}(r_{ij}-r_{ij}^*))  \right] \ ,
\label{Eq. Rupture_param}
\end{equation}
describes the effect of bond breakage. Here $r_{ij}^*=(R^{{\rm vdW}}_{ij}+r_0)/2$, with $r_0$ being the equilibrium distance between two atoms involved in the angular interaction and $R^{{\rm vdW}}_{ij}$ being the van der Waals radius for those atoms.

\begin{table*}[t!]
\centering
\caption{Parameters of the covalent bonded and angular interactions for the Pt(PF$_3$)$_4$ molecule employed in the present simulations.}
\begin{tabular}{p{3.5cm}p{1.5cm}p{1.5cm}|p{4.0cm}p{1.5cm}p{1.5cm}p{1.5cm}}
\hline
bond type      &   Pt--P     &   P--F         &  angle type  & P--Pt--P & Pt--P--F & F--P--F   \\
\hline
$r_0$~(\AA)      &    2.30     &     1.59    &  $\theta_0$~(deg.)      &  109.5   &  119.3 &  98.2 \\
$D_{ij}$~(kcal/mol) &    31.1     &    135.2    &       &     &  &  \\
$k_{ij}^{r}$~(kcal/mol \AA$^{-2}$) & 120.0 &   120.0   &  $k_{ijk}^{\rm {\theta}}$~(kcal/mol rad$^{-2}$)     &  76.4   &  28.0  &  100.0 \\
\hline
\end{tabular}
\label{Table:CovBonds}
\end{table*}

In the case study considered the initial geometry of a Pt(PF$_3$)$_4$ molecule is determined via the DFT calculation and then optimized using MBN Explorer. The rCHARMM parameters for a Pt(PF$_3$)$_4$ molecule are determined from a series of DFT-based potential energy scans, similar to how it was done in Ref.~\cite{DeVera2019} for a W(CO)$_6$ precursor molecule. In brief, the DFT calculations are performed using Gaussian 09 software \cite{Gaussian09} employing the B3LYP exchange-correlation functional and a mixed LanL2DZ/6-31+G(d,p) basis set, wherein the former set described the Pt atom and the latter is applied to P and F atoms. Geometry of the molecule is optimized first (states with spin multiplicity $M = 1, 3$ and 5 were considered) and a potential energy scan is then performed for Pt--P and P--F bonds to calculate equilibrium bond lengths, dissociation energies and force constants. The parameters of the bonded and angular interactions for Pt(PF$_3$)$_4$ are listed in Table~\ref{Table:CovBonds}. In contrast to the previous IDMD-based simulations \cite{Sushko2016, DeVera2020} of FEBID of W(CO)$_6$, angular interactions are explicitly included in the simulations reported here as they are crucial for describing correctly the tetrahedral structure of PF$_3$ ligands bound to the Pt atom.

\begin{figure}[t!]
\centering
\includegraphics[width=0.45\textwidth]{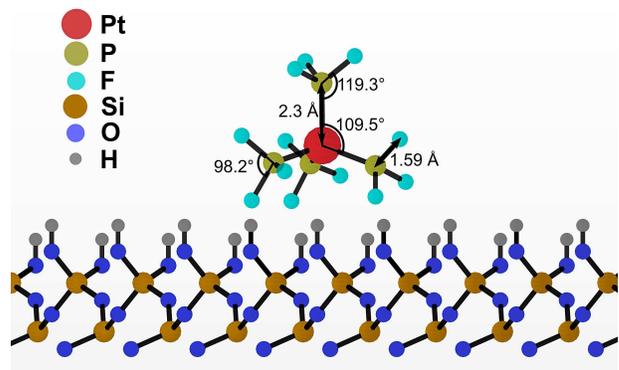}
\caption{Side view on an optimized single Pt(PF$_3$)$_4$ molecule adsorbed on a SiO$_2$-H substrate. The energy minimization calculation is performed using the interatomic potentials (\ref{Eq. Morse})--(\ref{Eq. Lennard-Jones}).}
\label{Fig:molecule}
\end{figure}

Following the earlier studies of FEBID by means of IDMD \cite{Sushko2016,DeVera2020}, the fully hydroxylated SiO$_2$ substrate is fixed in space in the course of simulations to speed up the simulations.
The interaction between the ideal SiO$_2$-H surface and adsorbed Pt-containing precursor molecules and fragments is governed by weak van der Waals bonding, which agrees with the results of Ref.~\cite{Plessow2016}.
Van der Waals forces between the atoms of the substrates and the adsorbed molecules are described by means of the Lennard-Jones potential:
\begin{equation}
    U_{{\rm LJ}}(r_{ij})=\varepsilon_{ij} \, \left [ \left (\frac{r^{{\rm min}}}{r_{ij}} \right )^{12}-2\left (\frac{r^{{\rm min}}}{r_{ij}}\right )^6 \right ]  ,
    \label{Eq. Lennard-Jones}
\end{equation}
where $\varepsilon_{ij}=\sqrt{\varepsilon_i \, \varepsilon_j}$ and $r^{{\rm min}} = (r^{{\rm min}}_i+r^{{\rm min}}_j)/2$.
Note that partial hydroxylation, surface defects and broken O-H bonds may lead to a stronger interaction between Pt and substrate atoms, causing agglomeration of Pt atoms at specific sites.
In this case the covalent interaction between Pt atoms and atoms of the SiO$_2$ can be described using the well-established interatomic potentials \cite{Plessow2016}. These effects can be addressed by the presented methodology in future studies.

The van der Waals parameters for all atoms in the system are taken from Refs.~\cite{Athanasopoulos1992, Mayo1990, Butenuth2012a} and are summarized in Table~\ref{Table:van_der_Waals}.
The parameters of the van der Waals interaction have been verified by simulating the surface diffusion of a single Pt(PF$_3$)$_4$ molecule on top of the fully hydroxylated SiO$_2$ layer. Although the surface diffusion data for Pt(PF$_3$)$_4$ on either hydroxylated or pristine SiO$_2$ are not available, the obtained value of the surface diffusion coefficient, $D = 2.85 \times 10^{-7}$~cm$^2$/s at 300~K, is within the range of values known for the typical surface diffusion coefficient of FEBID precursors \cite{Utke2008, Barth2020_JMaterChemC, Sushko2016}.
Figure~\ref{Fig:molecule} shows the optimized geometry of a Pt(PF$_3$)$_4$ molecule adsorbed on the SiO$_2$-H substrate and indicates the values of equilibrium covalent bond lengths and angles.

\begin{table}[t!]
\centering
\caption{Parameters of the Lennard-Jones potential describing the van der Waals interaction. $\varepsilon$ is the depth of the potential energy well and $r^{{\rm min}}$ is the interatomic distance corresponding to the potential energy minimum.}
\begin{tabular}{p{1.5cm}p{2.5cm}p{2.0cm}p{1.5cm}}
    \hline
	 Atom	& $\varepsilon$ (kcal/mol) & $r^{{\rm min}}/2$~(\AA) & Ref. \\
    \hline	
	Pt    & $15.72$  & 1.43  &  \cite{Athanasopoulos1992} \\
	P     & $0.32$   & 2.08  &  \cite{Mayo1990} \\
	F     & $0.07$   & 1.74  &  \cite{Mayo1990}    \\
	Si    & $0.30$   & 1.60  &  \cite{Butenuth2012a} \\
    O     & $0.26$   & 1.76  &  \cite{Butenuth2012a}  \\
	H     & $0.02$   & 1.00  &  \cite{Butenuth2012a}    \\
	\hline
\end{tabular}
\label{Table:van_der_Waals}
\end{table}

The interaction between metal atoms in the nanostructures formed on a surface can be described via common many-body potentials of the Embedded-Atom-Method type, e.g. Sutton-Chen \cite{SuttonChen1990} or Gupta \cite{Cleri1993a} potentials.
In this study the interaction between Pt atoms in the formed clusters is described by means of the Gupta potential \cite{Cleri1993a} with the cutoff distance of 7~\AA.

Atoms of both the precursor molecules and the substrate may carry non-zero partial charges, and thus the electrostatic interaction should also be taken into consideration. However, accounting for this interaction slows down the simulations significantly. In the present study, test simulations of the Pt(PF$_3$)$_4$ adsorption and diffusion on SiO$_2$-H with non-zero and zero partial charges on the Pt(PF$_3$)$_4$ molecules have been performed. On the basis of the results obtained, it was concluded that accounting for the Coulomb interaction makes a negligible difference in the structure of adsorbed molecules and their dynamics on the surface. Thus, in the following FEBID simulations the Coulomb interaction is omitted in order to improve the computational performance.

\subsection*{Step 2. Electron irradiation parameters}
\label{Method:Irradiation parameters}

Parameters of the electron interaction with a substrate and precursor molecules should be determined at Step~2 of the protocol shown in Figure~\ref{Fig:FEBID_protocol}.
First, one needs to determine the spatial distribution of secondary (SE) and backscattered (BSE) electrons produced due to the collision of the PE beam with the substrate. The convolution of the SE and BSE flux density with the fragmentation cross section of the precursor molecule determines the fragmentation probability of precursors at any space point within the system per primary electron. The electronic collisions with precursor molecules lead to their electronic excitation followed by the fragmentation and energy transfer to the recoil fragments. In the following subsections these processes are discussed in detail.

\begin{figure*}[t!]
\centering
\includegraphics[width=0.8\textwidth]{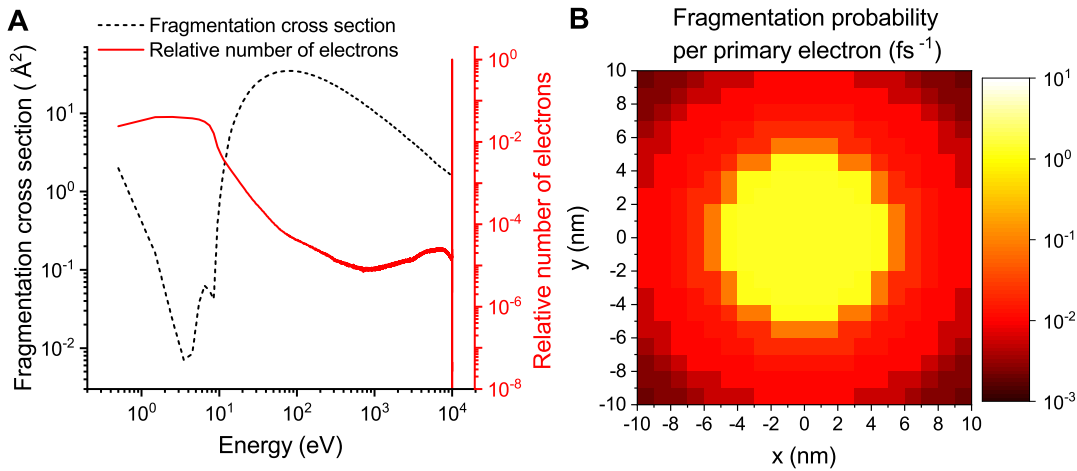}
	\caption{\textbf{A:} Total electron impact fragmentation cross section of Pt(PF$_3$)$_4$ including DEA and DI contributions (dashed line) and the relative number of electrons of specific energy per PE \cite{DeVera2020} including SE and BSE (solid line). \textbf{B:} Spatial distribution of the fragmentation probability of Pt(PF$_3$)$_4$ on SiO$_2$-H, irradiated with a 10~keV electron beam focused within the beam spot of 5~nm radius. The presented distribution is based on the defined PE flux of 1~\AA$^{-2}$fs$^{-1}$. }
\label{Fig:Cross-section}
\end{figure*}

\subsubsection*{PE, SE and BSE distributions}
The energy and the shape of the PE beam are used to define the distributions of SE and BSE. Typical PE energies used in the FEBID process vary from 1~keV to 30~keV \cite{Huth2012}. One can simulate a tightly-focused electron beam with different geometries (e.g. a cylindrical beam or a uniform broad beam covering the whole simulated surface). The latter case is relevant for surface science experiments on irradiation of thin films of adsorbed precursor molecules.

The yields of SE and BSE and the corresponding spatial distributions  depend on the energy of the PE beam and the material of a substrate. These distributions can be obtained by means of Monte Carlo (MC) simulations of electron transport \cite{Dapor_el-transport_2020}. There are several codes suitable for this purpose with different possibilities and limitations, for example, SEED \cite{Azzolini2019}, Geant4 \cite{Allison2016} and PENELOPE \cite{Baro1995-PENELOPE}. The yield of SE generated in various materials can also be evaluated by means of analytical models \cite{Lin_2005_SurfInterfAnal}.

The obtained spatial and energetic distributions of PE, SE and BSE are tabulated. The spatial distribution of electrons is defined on a cubic grid covering the whole simulation box; the grid consists of voxels with the size of 1~nm. The energy distribution of electrons is tabulated with a step of 1~eV. In order to permit a straightforward adjustment of the simulated electron flux, the electron distributions are scaled to a single primary electron per unit area within the irradiated surface area. At a small thickness of the deposited material, when the contribution of the deposit to the yield and distribution of electrons is negligible and the fragmentation process occurs mainly in the vicinity of the substrate surface, the radiation field can be defined once as a function of only the surface coordinates. In general, the electron distributions can be recalculated in the course of the FEBID process to account for electron interactions within the deposits. This adjustment is important for simulating the growth of three-dimensional (3D) structures.

In this study we employ the distribution of electrons calculated previously \cite{DeVera2020} using the SEED code and consider a cylindrical PE beam with a radius of 5~nm and energy of 10~keV. The number of generated electrons (that is the sum of SE and BSE contributions) of specific energy per primary electron \cite{DeVera2020} is shown in Figure~\ref{Fig:Cross-section}A by the solid red line. The dashed line shows the Pt(PF$_3$)$_4$ fragmentation cross section, discussed further in the text. 
Figure~\ref{Fig:Cross-section}B shows a spatial distribution of the fragmentation probability of Pt(PF$_3$)$_4$ per PE, calculated using the electron distributions and the Pt(PF$_3$)$_4$ fragmentation cross section.

\subsubsection*{Fragmentation cross section}

An overview of the electron interactions with precursor molecules including possible mechanisms of irradiation-induced molecular fragmentation can be found e.g. in Ref.~\cite{MooreH.2012}.
The main mechanisms of precursor fragmentation are the dissociative electron attachment (DEA) at low electron energies below the ionization potential of the molecule and the dissociative ionization (DI) at higher electron energies. The neutral dissociation (ND) is another possible fragmentation channel \cite{Zlatar2016}, but experimental data on ND are scarce.
Accounting for the ND fragmentation channel might lead to an increase of the fragmentation probability. This means that a given fragmentation probability will be achieved at slightly lower beam current. Such a behavior makes the FEBID outcomes not very sensitive to specific fragmentation channels and permits to compare the characteristics of the simulated FEBID process with experimentally observed characteristics at given experimental conditions.

Electron impact induced fragmentation experiments performed for a number of precursor molecules revealed \cite{Engmann2013, Thorman2015, Wnorowski2012} that the sum of partial cross sections of ionization resulting in emission of positive ion fragments exceeds significantly (by an order of magnitude) the cross section of ionization without fragmentation of the molecule.
On this basis one concludes that almost every ionizing collision leads to fragmentation; hence the DI cross section can be approximated by the total ionization cross section. The latter can be calculated by means of the dielectric formalism \cite{DeVera2013_PRL} as it was done in Ref.~\cite{DeVera2020} or by means of well-established semi-empirical methods. One of such methods is based on the additive rule \cite{DEUTSCH1989}, according to which the total molecular ionization cross section is approximated by the sum of ionization cross sections of the constituent atoms or smaller molecular fragments. This method was later improved by taking into account the molecular bonding \cite{DEUTSCH1994}. Ionization cross section can also be calculated using the Deutsch and M\"ark (DM) formalism \cite{Mark_1992} or the Binary Encounter Bethe (BEB) model \cite{Hwang1996}. It should be noted that the mentioned methods yield total electron ionization cross section of the target molecules, i.e. the sum of partial cross sections for all the channels leading to the formation of any charged molecule or fragment \cite{Huber2019}.

The IDMD approach permits considering several different fragmentation channels. The total fragmentation cross section of a bond is summed up from partial cross sections of DEA and DI channels leading to bond cleavage between atoms. The energy-resolved fragmentation cross section for a specific bond is used to calculate the space resolved fragmentation probability $P(x,y)$ per unit time \cite{DeVera2020}:
\begin{eqnarray}
    P(x,y) &=& \sigma_{\rm frag}(E_0) J_{\rm PE}(x,y,E_0) \nonumber \\
    &+& \sum_i \sigma_{\rm frag}(E_i) J_{\rm SE/BSE}(x,y,E_i)  \ .
\label{Eq. Frag_Probability_total}
\end{eqnarray}
Here $E_i < E_0$ is the electron energy discretized in steps of 1~eV, and $J_{\rm PE/SE/BSE}$ is the flux density of PE, SE and BSE, respectively.

The dependence of the fragmentation cross section of Pt(PF$_3$)$_4$ on electron energy, $\sigma_{\rm frag}(E)$, is shown in Fig.~\ref{Fig:Cross-section}A by the dashed black line. Here, the dissociation of Pt(PF$_3$)$_4$ is considered according to the reaction Pt(PF$_3$)$_4$ + $e^- \rightarrow$ Pt(PF$_3$)$_{4-n}^-$ + $n$PF$_3$. The electron-induced dissociation of P--F bonds has not been considered in the present simulations. The absolute cross section of DEA for Pt(PF$_3$)$_4$ is taken from the experiment \cite{May2012a}. The DI cross section is approximated from the total ionization cross section of the structurally similar molecule cisplatin (Pt(NH$_3$)$_2$Cl$_2$) \cite{Huber2019} by scaling the total number of electrons in the ligands. The accuracy of the cross section evaluation does not affect the results of FEBID simulations as a slight variation of the cross section (e.g. determined by means of more accurate theoretical or computational approaches) would correspond to slightly different values of beam current.

A product of space- and energy-resolved electron distribution and the precursor fragmentation cross section gives the fragmentation probability per PE per unit time.
The calculated probability based on the defined PE flux of 1~\AA$^{-2}$fs$^{-1}$ in the tabulated form for a 20~nm~$\times$~20~nm grid covering the simulation box (see Fig.~\ref{Fig:Cross-section}B) is used as input for the simulations of the irradiation phase of the FEBID process (see Step~4).

\subsubsection*{Energy transfer to the medium}

A projectile electron interacting with the precursor molecule transfers some amount of energy to the system leaving the molecule in an excited electronic state. In the case of ionization, some fraction of deposited energy is spent in overcoming the ionization threshold; another fraction is carried away by the ejected electron, while the remaining part is stored in the target in the form of electronic excitations. The latter can involve different molecular orbitals, being of either bonding or antibonding nature. An excitation involving an antibonding molecular orbital evolves through cleavage of a particular bond on the femtosecond timescale, and the excess energy is transferred to kinetic energy of the produced fragments.
The amount of energy transferred by the incident radiation to the system can be evaluated from quantum mechanical calculations of the processes of energy deposition and excitation \cite{Zlatar2016}. Experimentally, the transferred energy can be estimated based on the measured electron energy loss spectra for a specific precursor molecule \cite{May2012a}.

The amount of energy deposited locally into a specific covalent bond of the target and converted into kinetic energy of the two atoms forming the bond is as an input parameter for the simulations of FEBID, which may influence the rate of precursor molecule fragmentation. The chosen value of the deposited energy can be verified by comparing the dependence of surface coverage of precursor elements on the electron dose with the experimental data measured by means of X-ray photoelectron spectroscopy (XPS).

Electron energy loss spectra for Pt(PF$_3$)$_4$ molecules in the gas phase were measured experimentally and compared with time-dependent DFT (TDDFT) calculations in Ref.~\cite{Zlatar2016}.
In the simulations presented below two values of the energy deposited into the molecule, namely 205~kcal/mol (8.9~eV) and 300~kcal/mol (13~eV) were selected. These values correspond to positions of the peaks in the experimentally measured electron energy loss spectra for Pt(PF$_3$)$_4$ \cite{Zlatar2016}.

\subsection*{Step 3. Initial state of the system}
\label{Method:Initial_state}

In FEBID experiments precursor molecules are delivered by the gas injection system and fill the scanning electron microscope chamber. Depending on the experimental conditions (temperature, precursor gas pressure, geometry and position of the gas inlet), precursor molecules adsorb on the substrate surface and form a thin (from sub-monolayer to several monolayer thickness) film.
An explicit evaluation of adsorption and desorption rates for different substrates, temperatures and concentrations of the adsorbates can be performed by means of MD simulations. It should be stressed that, while the adsorption process can be simulated in detail on the atomistic level, the exact mechanism of precursor deposition does not affect the system's evolution during irradiation.
This means that only the final state of the molecular system, created before the irradiation, is essential for the IDMD simulations.

The outcome of atomistic simulations of precursor adsorption and desorption, i.e. the once defined regime, can be used for every related FEBID case study. Due to the complexity and multiple facets of the presented methodology, this question cannot be addressed in the present study and will be systematically explored in follow-up studies.

At Step~3 of the computational protocol (see Figure~\ref{Fig:FEBID_protocol}) a layer of precursor molecules of specific thickness and density is first created by means of the modeller plugin of MBN Studio \cite{Sushko2019} and then delivered to the substrate.
The molecular layer is created using the information on the geometry and topology of a single precursor molecule as follows. Precursor molecules are randomly distributed within a box positioned 3~nm above the substrate. Size of the box is defined according to the specified volume density of the molecules. The atomic coordinates and topology for all molecules in the created precursor layer are saved in .pdb and .psf formats, respectively. Size and molecular density of the precursor layer can be varied to adjust coverage of the substrate surface with precursor molecules corresponding to the experimental conditions.
In order to avoid non-physical overlapping of the atoms the layer is optimized first using the velocity quenching algorithm for 20,000 steps with the velocity quenching time step of 0.5~fs. After that the precursor molecules are pulled down to the surface and thermalized at the specified temperature to reach thermal equilibrium. In the simulations the substrate atoms are frozen while all the atoms in precursor molecules are freely moving in space.

In this study a sub-monolayer of Pt(PF$_3$)$_4$ with the size 20~nm$\times$~20~nm and thickness of 10~nm is optimized, adsorbed on SiO$_2$-H substrate and thermalized at 300~K for 0.1~ns using the Langevin thermostat with a damping time of 0.2~ps. The constructed layer consists of approximately 370 molecules that corresponds to surface density of 0.9 molecules per nm$^2$. These and all the subsequent simulations described further are performed using the Verlet integration algorithm with a time step of 1~fs and reflective boundary conditions. Interatomic interactions are treated using the linked cell algorithm \cite{Solovyov2012} with a cell size of 10~\AA.

\subsection*{Step 4. FEBID process}
\label{Method:FEBID_simulation}

The FEBID process (Step 4 in Figure~\ref{Fig:FEBID_protocol}) consists of two phases, namely irradiation and replenishment of the precursors, which are repeated multiple times.
The outcome of the FEBID process is governed by a balance of several processes, such as adsorption, desorption, diffusion and dissociation of molecules. The cumulative contribution of these processes defines the regime at which the FEBID goes, i.e. time dependent concentration and the distribution of precursors. Our approach is to replicate the regime representing the system’s state before, during and after irradiation, without necessarily explicit simulation of all the involved processes. The present methodology allows the inclusion of additional processes, such as electron stimulated desorption and chemical reaction between fragments, deposits and the substrate. Investigating the role of these processes at specific conditions might be a topic for a separate investigation, which goes beyond the scope of the present study.

\subsubsection*{Irradiation}
In experiments the irradiation phase of the FEBID process can last for a period (called dwell time $\tau_d$) from sub-microseconds to sub-milliseconds \cite{Utke2008}.
During the irradiation phase, the dissociation of the metal--ligand bonds in the precursor molecules occurs resulting in the emergence of reactive molecular fragments with dangling bonds. The dynamics and interaction of the fragments lead to the formation of new bonds between Pt atoms and their coalescence into metal-enriched clusters and bigger structures.
As the typical dwell time values are relatively short, adsorbtion and desorbtion of precursor molecules within the irradiation phase is assumed negligible.

At this step the electron induced precursor fragmentation is simulated by means of IDMD \cite{Sushko2016} using space-dependent bond dissociation rates for molecules on the substrate. In brief, these rates depend, in steady-state conditions, on (i) the number and energy of electrons crossing the substrate surface at each point per unit time and unit area (which in turn are determined by the PE beam energy $E_0$ and flux $J_0$), and (ii) the energy-dependent molecular fragmentation cross section $\sigma_{{\rm frag}}(E)$.

As described above at Step~2, the tabulated space-resolved fragmentation probability per primary electron is used in the IDMD simulation of the irradiation phase to link the bond dissociation rate to the electron flux. As the realistic experimental time scale for $\tau_d$ is challenging for all-atom MD, the simulated PE fluxes $J_0$ (and hence PE beam currents $I_0$) are rescaled to match the number of PE per unit area and per dwell time as in experiments. The correspondence of simulated results to experimental ones is established through the correspondence of the electron fluence per dwell time per unit area in simulations and experiments \cite{Sushko2016}. Such an approach is valid in the case when different fragmentation events occur independently and do not induce a collective effect within the system. In this case the irradiation conditions for the adsorbed precursor molecules are the same in simulations and in experiments. This correspondence condition gives
\begin{equation}
    I_{{\rm exp}} = I_{{\rm sim}} \, \frac{S_{{\rm exp}}}{\lambda S_{{\rm sim}} }= I_{{\rm sim}} \, \frac{R_{{\rm exp}}^2}{\lambda R_{{\rm sim}}^2 } \ ,
\end{equation}
\begin{equation}
\lambda = \frac{\tau_d^{{\rm exp}}}{\tau_d^{{\rm sim}}} \ ,
\end{equation}
where $S_{{\rm exp}}$ and $S_{{\rm sim}}$ are the electron beam cross sections used in experiments and simulations, respectively; $R_{{\rm exp}}$ and $R_{{\rm sim}}$ are the corresponding beam spot radii.

The following experimental irradiation parameters \cite{Barry2006a} have been used in the simulations performed in the present study:  electron current $I_{{\rm exp}} = 2.8$~nA and the estimated beam spot radius $R_{{\rm exp}} = 40$~nm. Although in the experiments \cite{Barry2006a} irradiation was performed with a continuous electron beam, we consider dwell time of a single irradiation cycle to be $\tau_d^{{\rm exp}}$~=~1~ms. This corresponds to the PE flux $J_0^{{\rm exp}} \approx 3480$~nm$^{-2}$ms$^{-1}$. Following the earlier studies of FEBID using IDMD \cite{Sushko2016, DeVera2020} the dwell time value $\tau_d^{\rm sim} = 10$~ns is used in the simulations. The rescaled electron current in the simulations is thus $I_{{\rm sim}} \approx 4$~$\mu$A.

\subsubsection*{Replenishment}

During the replenishment phase precursor molecules are not exposed to the electron beam, while volatile fragments leave the surface and new precursor molecules are adsorbed. The characteristic time of the replenishment phase in FEBID experiments is on the order of milliseconds \cite{Huth2012}. The amount of the added precursor molecules in combination with the electron current density defines whether the FEBID process runs in the electron-limited regime or the molecule-limited regime \cite{Barth2020_JMaterChemC}. The electron-limited regime corresponds to a large number of precursor molecules with a small number of electrons leading to slow fragmentation and accumulation of the residual precursors. In the case of the molecule-limited regime, the number of secondary electrons exceeds the number of adsorbed precursors, causing dissociation of a larger number of molecules.

The realistic experimental time scale for the replenishment phase is challenging for MD simulations as it would require simulating the slow adsorption and desorption of new precursors and the created molecular fragments from the surface occurring on the relatively large time scales.
For specific values of pressure and temperature, the duration of the replenishment phase defines the concentration of adsorbates and the amount of desorbed molecules at the beginning of the next irradiation phase. If the replenishment time is long enough, the steady state is achieved. For shorter replenishment times, the concentration of adsorbates will be a fraction of the steady-state concentration.

Similarly to the creation of the initial precursor layer (see Step~3), the replenishment phase is simulated to reproduce the physical state of the system after the replenishment, which is  characterized by a certain number of desorbed fragments and the spatial distribution of newly adsorbed precursor molecules. The amount and spatial distribution of the precursor molecules added at each replenishment step can be varied in the model to describe different experimental conditions. The spatial distribution of the adsorbates added within the replenishment phase can also be modified depending on the injection method of precursor gas.
The weakly bound precursor molecules and fragments are removed from the system by an artificial external force field. This procedure can be adjusted to the experimental desorption rate.
After that, a new layer of precursors is created at a certain distance above the surface, optimized and deposited upon the substrate.
Thermalization and relaxation of the system are performed at the end of the replenishment stage, which is then followed by the next cycle of irradiation.

In the present study each replenishment phase is simulated for 0.4~ns. New precursor molecules are deposited over the area of 14~nm~$\times$~14~nm to cover the beam spot while preventing the accumulation of the non-fragmented molecules along the perimeter of the simulation box where the fragmentation probability is very low (see Fig.~\ref{Fig:Cross-section}B). The density of new molecular layers is adjusted to maintain a balance between the number of added molecules and the number of fragmented molecules during the irradiation step.
The surface density of the added Pt(PF$_3$)$_4$ layer is 0.3 and 1.2 molecules per nm$^2$ for $E_{{\rm dep}}=205$~kcal/mol and $E_{{\rm dep}}=300$~kcal/mol, respectively. The deposition and thermalization simulations for each new layer are performed in the same way as for the initial precursor layer.

\subsection*{Step 5. Post-deposition processing}
\label{Method:Postprocessing}

Additional processing of the FEBID-grown structures, namely post-deposition electron irradiation or high-temperature annealing with or without additional gas (e.g. O$_2$, H$_2$O) \cite{Plank_2013_ACS_ApplMater, Shawrav2016a}, is commonly employed for the deponat purification purposes. According to the experimental data \cite{Botman2009a, Huth2018} as-grown FEBID structures usually contain a relatively low ($\sim 10 - 30\%$) amount of metal, whereas post-irradiation processing of the deposits enables to significantly increase the metal content up to 80\%.
Energy transferred into the system by heating or via the electron-molecule interaction activates desorption of volatile molecules from the surface and leads to rearrangement of the deposited structures. As a result, residual precursor molecules and fragments become removed, and remaining metal deposits are reorganized into more compact and dense structures.

Both types of post-deposition processing, i.e. annealing and further electron irradiation, can be simulated by means of MD as the last step of the computational protocol shown in Figure~\ref{Fig:FEBID_protocol}.

MD simulation of the annealing includes heating of the deposited structures to high temperature and subsequent slow cooling. During annealing structural and topological changes in the deposit take place.
The typical time scale for annealing in experiments varies from minutes to several hours \cite{Huth2018, dosSantos_2018, Che2005_APL}. Therefore the temperature and the duration of the annealing in the simulations should be rescaled to match the amount of heat transferred to the deposited structures in experiments.

The simulation of the post-growth electron irradiation without depositing new precursors should be performed in a similar way as Step~4, but without the replenishment phase.
This process is characterized by space- and energy-resolved distribution of the electrons, electron flux and duration of the irradiation. If the deposited structure is relatively thin (e.g. of a sub-monolayer thickness), in the first approximation one can use the same electron distribution as for the FEBID irradiation step. Otherwise, the SE distribution should be recalculated to account for the electron transport in both the substrate and the deposited material.

The computational procedure for atomistic modeling of the post-deposition effects has been included in the presented methodology shown in Figure~\ref{Fig:FEBID_protocol} for the sake of completeness.
Several test simulations of post-annealing and irradiation have been performed to verify the method, but a systematic analysis of these effects goes beyond the scope of the present paper and will be a subject of a separate study.

\section{Results and discussions}
\label{Results}

The above-described protocol for atomistic IDMD simulations of the FEBID process using the MBN Explorer and MBN Studio software has been used to simulate the formation of Pt-containing nanostructures during the FEBID of Pt(PF$_3$)$_4$ molecules. The results of these simulations are described in this section. Particular focus is made on the detailed structural analysis of nucleated metal clusters.

\begin{figure}[t!]
\centering
\includegraphics[width=0.45\textwidth]{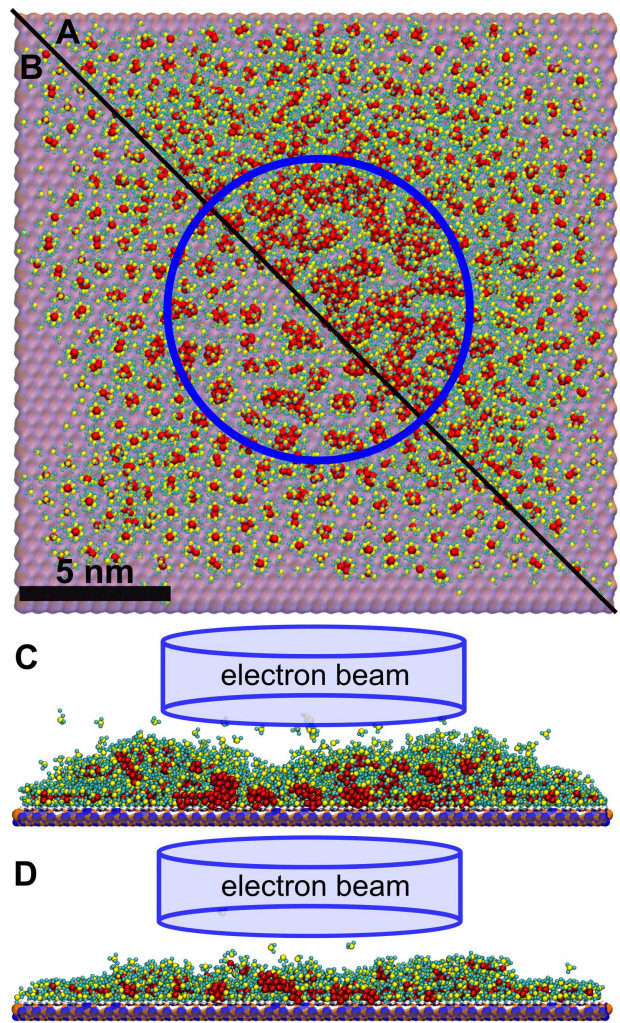}
	\caption{\label{Fig:Snapshot}A snapshot of the multiscale IDMD simulation of the FEBID process of Pt(PF$_3$)$_4$ at the end of the 20th cycle (after 200~ns of the simulation). Panels A and C show top view and side view of a 6 nm thick slice through the beam center of the grown structure for the regime of the energy transferred to the medium during fragmentation $E_{\rm dep} = 300$~kcal/mol with 223 precursor molecules added on average per FEBID cycle. Panels B and D show the corresponding views for the the regime of $E_{\rm dep} = 205$~kcal/mol with 68 precursor molecules added on average per FEBID cycle. The scale bar is the same for all figures. The blue cylinder indicates the PE beam spot. The image is rendered by means of VMD \cite{HUMP96}. }
\end{figure}

IDMD simulations have been performed for 30 cycles of the FEBID process with a total simulation time of 300~ns. The accumulated fluence of PE is $\sim 1.04 \times 10^{19}$~cm$^{-2}$, which corresponds to the equivalent total experimental irradiation time of 30~ms with the beam parameters from Ref.~\cite{Barry2006a}.
A snapshot of the system at the end of the 20th FEBID cycle is presented in Figure~\ref{Fig:Snapshot}. Panels~A--D correspond to different amounts of energy transferred to the medium via the Pt--P bonds fragmentation. The energy parameter $E_{\rm dep}$ governing the bond fragmentation is equal to 300~kcal/mol and 205~kcal/mol for the cases presented in panels A \& C and B \& D, correspondingly.
As shown in Figs.~\ref{Fig:Snapshot}A and \ref{Fig:Snapshot}B, three spatial regions can be distinguished where different structures are formed on the surface depending on the spatial distributions of the fragmentation probability and adsorbed precursor molecules. Inside the beam spot area with the diameter of 10~nm (indicated by the blue circle in Figs.~\ref{Fig:Snapshot}A and \ref{Fig:Snapshot}B) high probability of Pt--P bond fragmentation leads to dissociation of Pt(PF$_3$)$_4$ molecules and formation of metal clusters. The clusters grow, merge and interconnect during the irradiation process, forming a network of thread-like metallic nanostructures. The transition region of 1~nm radius outside the beam spot area contains smaller metal clusters with a larger number of PF$_3$ ligands attached due to lower fragmentation probability in this spatial region. The presence of the ligands prevents dense packing and aggregation of isolated metal clusters. As a result, the height of the deposited structures in this spatial region is higher than that within the beam spot area. The region beyond the transition region (near the simulation box boundaries) contains mostly intact or less fragmented precursor molecules.

The growth of Pt-containing clusters occurs faster with an increase of $E_{\rm dep}$. Within the $E_{\rm dep}$ energy range considered, the fragmentation of Pt(PF$_3$)$_4$ molecules is the most prominent at $E_{\rm dep} = 300$~kcal/mol (see Fig.~\ref{Fig:Snapshot}A and \ref{Fig:Snapshot}C). This can be explained by lowering the probability of rejoining the broken bonds at the higher amount of energy deposited therein. The number of precursor molecules added at each FEBID cycle is approximately equal to the number of fragmented molecules. The average number of added molecules is equal to 68 and 223 for $E_{\rm dep} = 205$~kcal/mol and $E_{\rm dep} = 300$~kcal/mol, respectively. These numbers correspond to the surface density of 0.3 and 1.2 molecules/nm$^2$, respectively.
The larger amount of precursors adsorbed on the surface in the FEBID process leads to faster accumulation of Pt atoms enabling the formation of larger metal clusters.

\begin{figure*}[t!]
\centering
\includegraphics[width=0.7\textwidth]{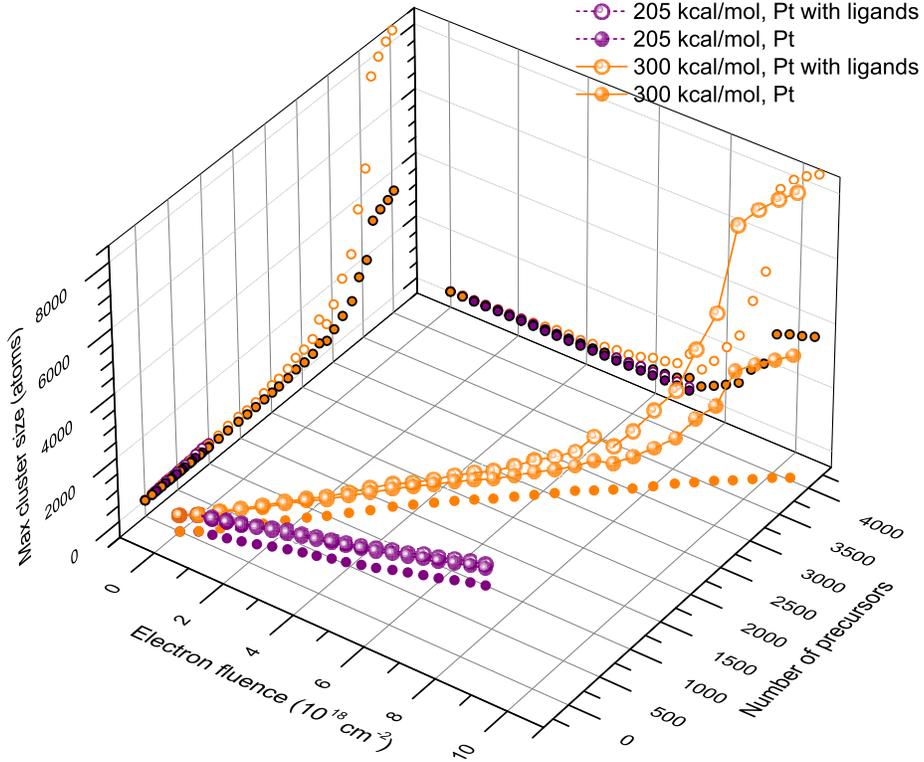}
	\caption{The number of Pt atoms (full spheres) and the total number of atoms including ligands (open spheres) in the largest cluster as a function of electron fluence and the number of adsorbed precursor molecules in the effective beam spot for the energy transferred to the medium during fragmentation $E_{\rm dep} = 205$~kcal/mol (purple symbols) and 300~kcal/mol (orange symbols). Full and open dots show the respective projections on the planes. Each symbol describes the parameter values derived at the end of each consecutive FEBID cycle.}
\label{Fig:Atom_Size}
\end{figure*}

The growth of the deposits can be quantified by the total number of atoms and the number of metal atoms in the largest cluster. Figure~\ref{Fig:Atom_Size} shows the evolution of maximal cluster size for $E_{\rm dep} = 205$ and 300~kcal/mol as a function of electron fluence and the number of Pt(PF$_3$)$_4$ molecules deposited within the beam spot area (with the radius of 5~nm) surrounded by a diffusive halo with the width of 1~nm. This spatial region (denoted hereafter as the effective beam spot area) is characterized by the highest fragmentation probability (see Fig.~\ref{Fig:Cross-section}B) due to secondary and backscattered electrons.
Figure~\ref{Fig:Atom_Size} illustrates that several stages of the nanostructure growth can be distinguished. For the case of $E_{\rm dep} = 300$~kcal/mol, fragmentation of the precursor molecules leads to the formation of isolated small clusters during the first 5-6 FEBID cycles (corresponding to the electron fluence of $\sim 2 \times 10^{18}$~cm$^{-2}$). This regime is characterized by a linear dependence of the largest structure size on electron fluence (see orange dots in Fig.~\ref{Fig:Atom_Size}). During further irradiation the clusters start to merge while the structure growth continues simultaneously due to the deposition of new precursor molecules. An interplay of these phenomena results in a much faster increase of the number of atoms in the largest cluster as a function of electron fluence.
The growth of Pt clusters as a function of the number of added precursor molecules follows the same trend for both values of $E_{\rm dep}$, but the growth rate is higher at the larger $E_{\rm dep}$. Simultaneously, much faster growth rate as a function of electron fluence is observed for $E_{\rm dep} = 300$~kcal/mol. The similar evolution of the maximum cluster size for both values of $E_{\rm dep}$ suggests that the results obtained in the two studied regimes can be scaled.

The maximum size of the metal cluster, $N_{\rm max}$, is determined by the number of metal atoms located in the effective beam spot area. The latter number depends, in turn, on the number of precursor molecules, $N_{\rm p}$, adsorbed in that spatial region at given irradiation conditions. The dependence of $N_{\rm max}$ on $N_{\rm p}$ follows the empirical dependence represented by the two terms corresponding to the two regimes of the cluster growth described above:
\begin{equation}
    N_{\rm max} = a N_{\rm p} + N_{\rm lim} \left[ 1+e^{- \frac{N_{\rm p} - N_{\rm tr}}{\Delta N}} \right]^{-1} \ .
\label{Eq. MaxClusterSize1}
\end{equation}
Here $a$ is the dimensionless coefficient of the initial linear growth of clusters. The beam spot size limits the maximal transversal size of the formed metal nanostructure. This condition is accounted for by introducing a sigmoid function, the second term on the right-hand side of Eq.~(\ref{Eq. MaxClusterSize1}). $\Delta N$ is an interval of $N_{\rm p}$ values within which the morphological transition from isolated islands to a single nanostructure takes place. The parameter $N_{\rm lim}$ stands for the limiting cluster size at which the morphological transition will be completed. $N_{\rm tr}$ is the sigmoid's midpoint that defines the number of adsorbed precursor molecules enabling the morphological transition. In general, all these parameters depend on the irradiation and the replenishment conditions as well as on the value of $E_{\rm dep}$.

\begin{figure}[t!]
\centering
\includegraphics[width=0.48\textwidth]{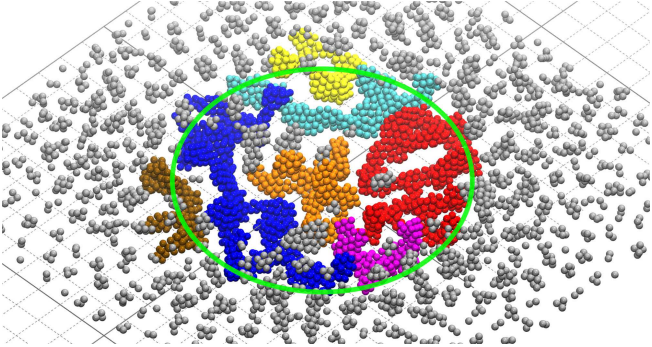}
	\caption{A snapshot of the IDMD simulation of the FEBID process of Pt(PF$_3$)$_4$  in the regime of $E_{\rm dep} = 300$~kcal/mol at electron fluence of $8 \times 10^{18}$ cm$^{-2}$. Only Pt atoms are shown. Separate clusters of a size larger than 150 atoms are shown in different colors, while smaller clusters are shown in grey. The green circle indicates the PE beam spot. The grid spacing is 1~nm.}
\label{Fig:Pt_cluster_Snapshot}
\end{figure}

Equation~(\ref{Eq. MaxClusterSize1}) permits the quantitative description of the morphological transition from separate clusters into a single structure.
As shown previously \cite{Sushko2016}, when isolated metal islands merge into a single nanostructure, its further growth occurs due to the attachment of fragmented precursor molecules to it. In this regime the maximal size of the metal structure follows a linear dependence on electron fluence and the number of precursor molecules.
The parameters of the sigmoid function can be defined with a high precision at least after passing the transition point. As shown in Figure~\ref{Fig:Pt_cluster_Snapshot}, several large clusters containing more than 150 Pt atoms have been formed in the beam spot area in the present simulation at electron fluence of $8 \times 10^{18}$~cm$^{-2}$. Within the next several cycles these clusters merge together and the structure starts to grow by the attachment of new material to the existing deposit. This effect is characterized as the change of the growth rate during the last four simulation cycles in Fig.~\ref{Fig:Atom_Size}.

The number of precursor molecules added at each FEBID cycle depends on the value of $E_{\rm dep}$.  As seen from Figure~\ref{Fig:Atom_Size}, the dependence of $N_{\rm p}$ on electron fluence $F$ can be approximated by a linear function:
\begin{equation}
    N_{\rm p} = k (E_{\rm dep}) \cdot S_0 \cdot F \ ,
\label{Eq. NumberOfMolecules}
\end{equation}
where  $S_0 = 1.1 \times 10^{-16}$~cm$^2$ is the area of the spherical region with the radius of 6~nm. 
Equation~(\ref{Eq. NumberOfMolecules}) is valid for the case when the amount of adsorbates added at each sequential FEBID cycle is kept nearly constant. In the present case, this amount is defined by the number of fragmented precursors. The ratio of newly delivered and fragmented precursors defines whether the considered regime is precursor-limited or electron-limited.
Fitting the data plotted in Fig.~\ref{Fig:Atom_Size} yields the values $k_{205} = 1.3 \times 10^{-4}$ and $k_{300} = 3.6 \times 10^{-4}$ for $E_{\rm dep} = 205$ and 300~kcal/mol, respectively.
Substituting Eq.~(\ref{Eq. NumberOfMolecules}) in Eq.~(\ref{Eq. MaxClusterSize1}), one derives the dependence of the maximal cluster size on electron fluence:
\begin{equation}
    N_{\rm max} = a k(E_{\rm dep}) \, S_0 F + N_{\rm lim} \left[ 1+e^{- \frac{k S_0 F - N_{\rm tr}}{\Delta N}} \right]^{-1} \ .
\label{Eq. MaxClusterSize2}
\end{equation}
Equation~(\ref{Eq. MaxClusterSize2}) has been used to fit the simulated dependencies $N_{\rm max}(F)$ for $E_{\rm dep} = 205$ and 300~kcal/mol with the variable parameters $a$, $\Delta N$, $N_{\rm tr}$ and $N_{\rm lim}$, whereas the values of $k$ have been fixed. As the dependence of $N_{\rm max}$ on $N_{\rm p}$ is practically the same at the different $E_{\rm dep}$ (see Fig.~\ref{Fig:Atom_Size}), the variable parameters are set the same for both dependencies. The fitting procedure yields the values summarized in Table \ref{Table:ClusterSizeFit}. The successfully converged fit confirms that the variable parameters defining the growth rate of the clusters are independent on $E_{\rm dep}$.

\begin{table}[t!]
	\caption{\label{Table:ClusterSizeFit} Parameters from Eq.~(\ref{Eq. MaxClusterSize2}) providing the best fit of the analytical dependence with the one obtained from simulations. $k$ is a scaling coefficient of the number of adsorbed precursors, $S_0$ is the effective beam spot area, $N_{\rm lim}$ is the estimated limiting number of Pt atoms in the nanostructure, $N_{\rm tr}$ is the number of adsorbed precursors in the transition point, $\Delta N$ is an interval of $N_{\rm p}$ values within which the morphological transition takes place, $a$ is the coefficient of the linear cluster growth. }
	\centering
	\begin{tabular}{p{1.6cm}|p{2.2cm}p{2.2cm}}
	Parameter	& \multicolumn{2}{c}{$E_{\rm dep}$} \\
        \hline
	         	& 205~kcal/mol  & 300~kcal/mol  \\
		\hline
	$k$             & $1.3 \times 10^{-4}$   &$3.6 \times 10^{-4}$  \\
	$S_0$, cm$^2$   & \multicolumn{2}{c}{$1.1 \times 10^{-16}$ } \\
	$N_{\rm lim}$   & \multicolumn{2}{c}{3174} \\
	$N_{\rm tr}$        & \multicolumn{2}{c}{3544} \\
	$\Delta N$      & \multicolumn{2}{c}{181}    \\
	$a$             & \multicolumn{2}{c}{$8.1 \times 10^{-2}$}   \\
	\hline
	\end{tabular}
\end{table}

The similar $N_{\rm max}(N_{\rm p})$ dependence at different $E_{\rm dep}$ values permits to explore computationally this dependence at larger $E_{\rm dep}$ (as such simulations are significantly faster) and then rescale it to the physically meaningful, smaller value of $E_{\rm dep}$ and larger fluence $F$ according to Eq.~(\ref{Eq. NumberOfMolecules}). One should note that the similar $N_{\rm max}(N_{\rm p})$ dependencies have been observed when the number of precursor molecules added at each FEBID cycle is approximately equal to the number of fragmented molecules for both values of $E_{\rm dep}$. The possibility of scaling for different FEBID regimes will be explored in future studies.

\begin{figure*}[t!]
\centering
\includegraphics[width=0.7\textwidth]{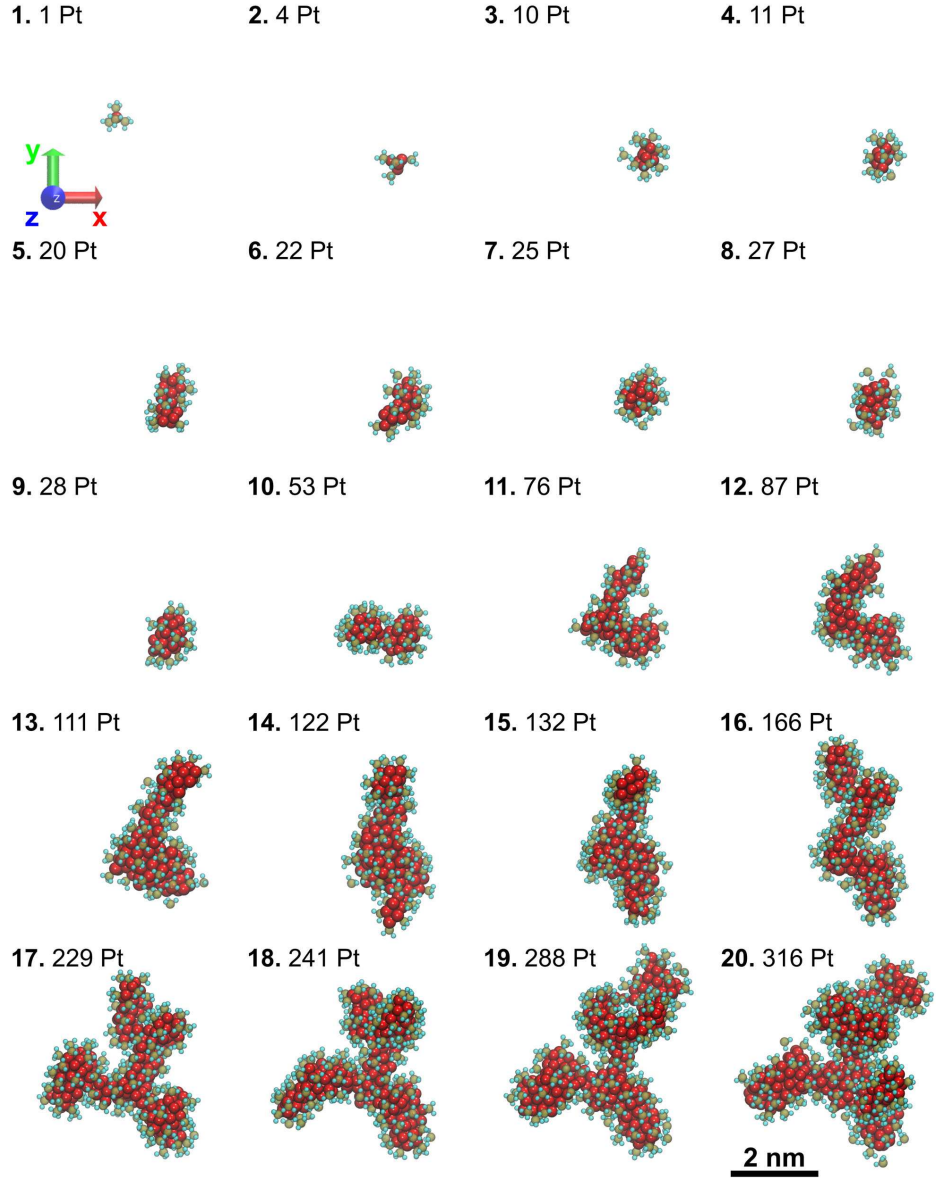}
	\caption{Evolution of the largest cluster in the course of simulation at the end of each FEBID simulation cycle (shown in bold). The number of Pt atoms is also indicated for each structure.}
\label{Fig:Cluster_Growth}
\end{figure*}

In the following subsections the deposited nanostructures are characterized in greater detail by considering the evolution of (i) the individual cluster size and shape, and (ii) height and metal content of the whole deposited structure in the beam spot area as a function of electron fluence or the number of irradiation cycles. The results presented below correspond to the case of $E_{\rm dep} = 300$~kcal/mol.

\subsection{Characterization of individual clusters}

One of the largest clusters formed in the course of the simulation has been selected to study its evolution during the FEBID process. The cluster's evolution is tracked back to the initial nucleation stage using coordinates of the center of mass of the metal core. The cluster structure and the number of Pt atoms in the cluster at the end of each FEBID cycle are shown in Figure~\ref{Fig:Cluster_Growth}. The platinum-containing nanostructure grows via coalescence of the neighboring metal clusters of different sizes. The coalescence takes place via an interplay of the following mechanisms:
an addition of a fragmented precursor molecule with a single Pt atom (see the evolution of the cluster structure at FEBID cycles 3-4 and 8-9), merging of two clusters of the comparable sizes (see the cluster structure at the cycles 4-5, 9-10 and 10-11), as well as a combination of both.
Metal clusters containing up to about 30 Pt atoms preserve a spherical shape and tend to rearrange after an elongation caused by the merging of clusters of comparable size (see cycles 1--9). The sequential coalescence of larger clusters containing several tens of Pt atoms results in the formation of randomly oriented branched structures (see cycles 10--20). The continuation of the irradiation and replenishment processes leads to further growth and interconnection of the branched clusters into a single metal network observed in this simulation at 27th cycle corresponding to PE fluence of 9.4 $\times 10^{18}$~cm$^{-2}$ (see Figure~\ref{Fig:Atom_Size}).

\begin{figure}[t!]
\hspace{-1.0cm}
\includegraphics[width=0.53\textwidth]{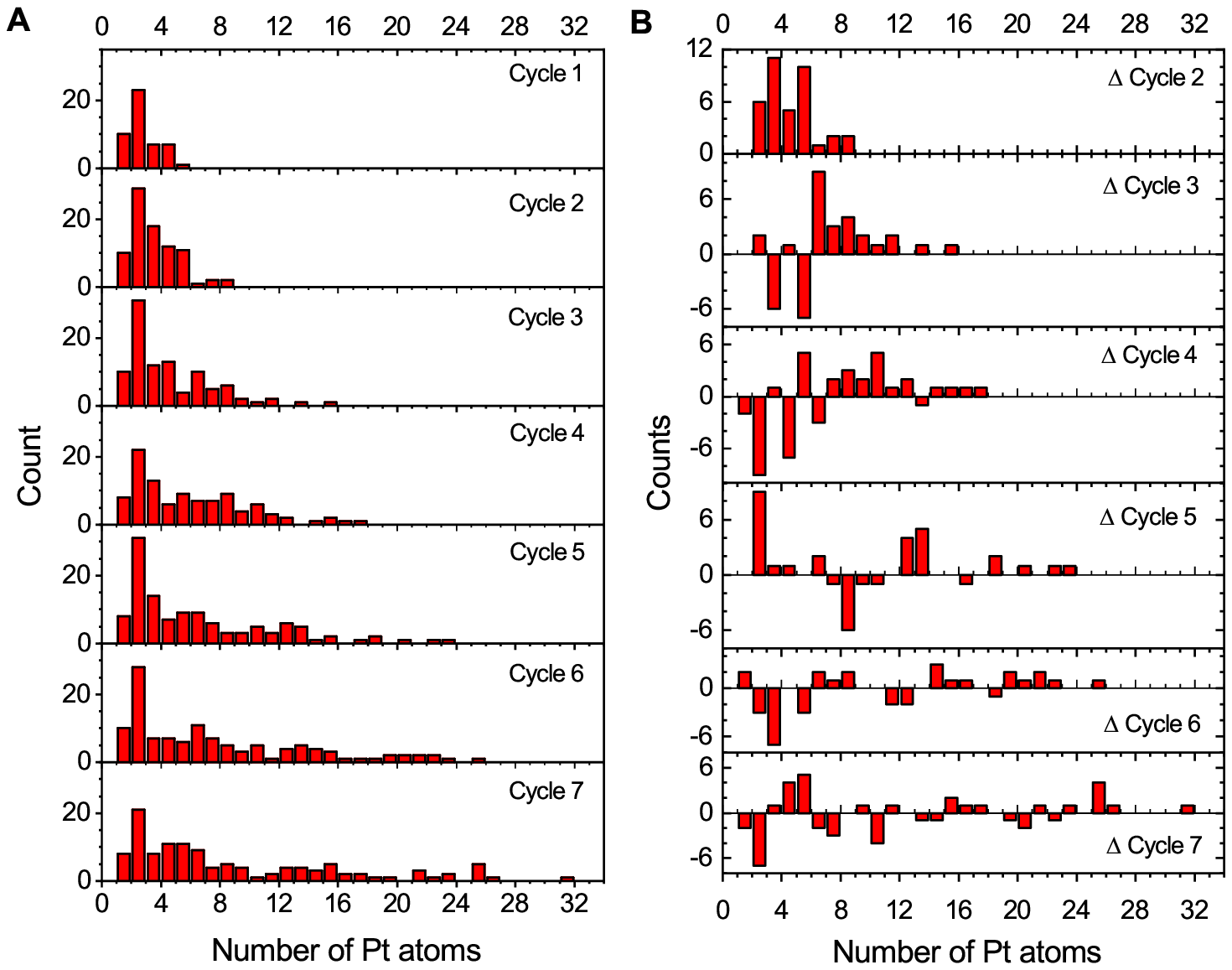}
	\caption{A) Size distribution of Pt clusters during the first 7 FEBID cycles.
	B) Difference of size distributions for Pt clusters between two consecutive FEBID cycles for the first 7 cycles. }
\label{Fig:AtomsSizeDistribution}	
\end{figure}

Next, the process of initial cluster growth and coalescence has been studied for an ensemble of the deposited clusters.
Figure~\ref{Fig:AtomsSizeDistribution} shows the size distribution of grown clusters as a function of the number of Pt atoms
(panel~A) and the difference between the distributions for two consecutive FEBID cycles (panel~B) for the first 7 FEBID cycles. In the course of these cycles small metal clusters start to nucleate and reach the size of about 20-30 atoms. As demonstrated in Fig.~\ref{Fig:Cluster_Growth}, clusters of substantially larger size are formed during the several follow-up FEBID cycles by merging the clusters of similar size, which have been formed over the first 7 cycles. The distributions shown in Fig. \ref{Fig:AtomsSizeDistribution} are obtained at the end of the irradiation stage at each FEBID cycle. Positive values in Fig.~\ref{Fig:AtomsSizeDistribution}B indicate an increased number of clusters of a given size in comparison with the previous cycle; negative values indicate a decreased number of the clusters of such size. As more irradiation cycles are performed and more Pt atoms are accumulated on the surface due to replenished precursors, the Pt-containing structures start to merge and consistently increase in size. The size distributions shown in Fig.~\ref{Fig:AtomsSizeDistribution}A are peaked at the number of platinum atoms $N = 2$ for all the FEBID cycles considered. This feature is attributed to the constant addition of precursor molecules during the replenishment stage at each FEBID cycle. During the following cycles Pt-containing structures are formed mainly via coalescence of larger clusters containing about $20-30$ Pt atoms. When the clusters reach a certain size, they become less mobile and behave as centers of attraction for new molecules.

\begin{figure}[t!]
\centering
\includegraphics[width=0.48\textwidth]{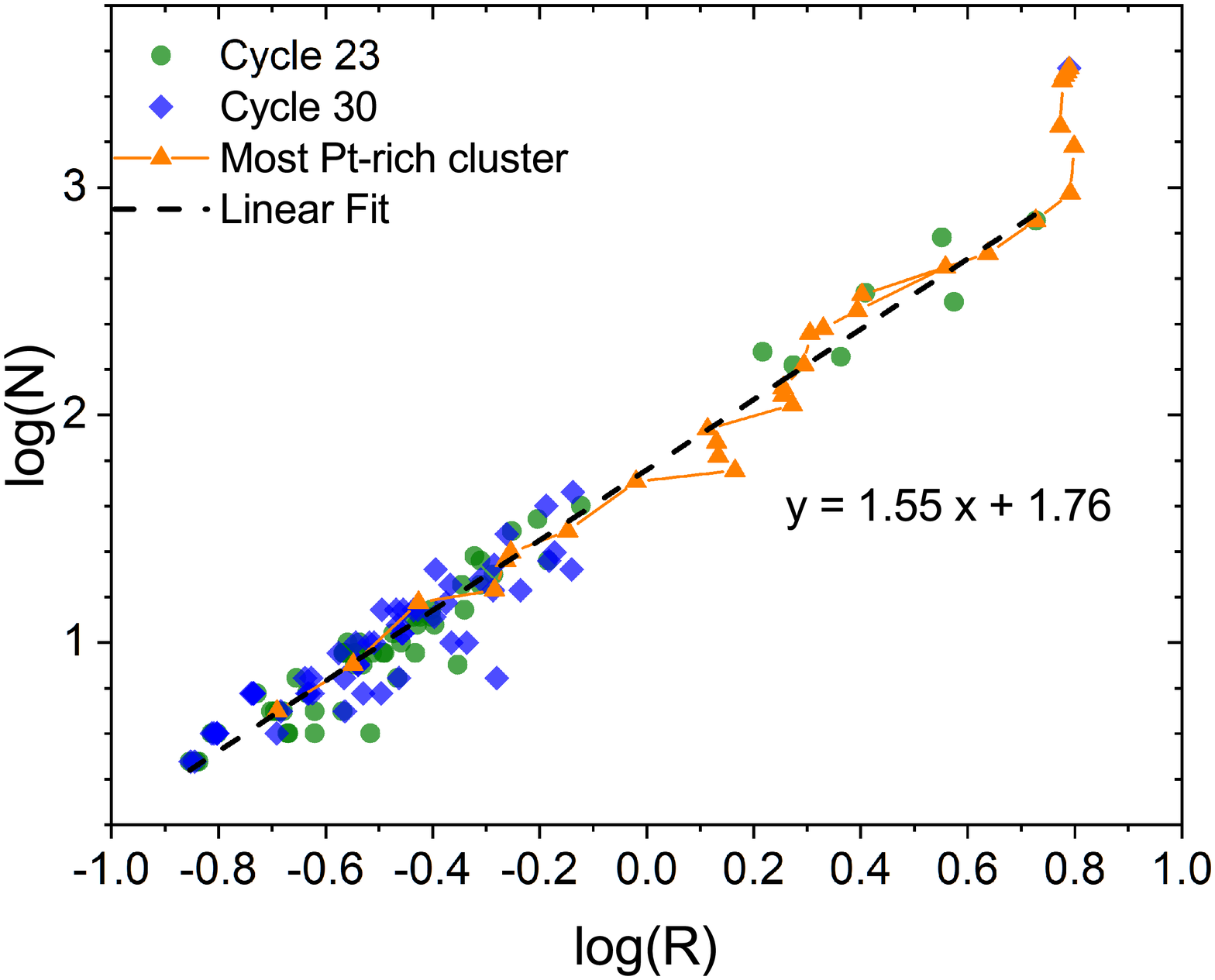}
	\caption{\label{Fig:FractalDimension} Double-log plot of the number of Pt atoms $N$ in the deposited clusters vs. the radius $R$ of the circumscribed sphere. Green dots and blue diamonds correspond to the whole ensemble of deposits at the end of the 23rd and 30th simulation cycle, respectively. Interconnected orange triangles describe the cycle-by-cycle growth of the most Pt-rich clusters. The dashed line shows a linear fit of the $\log{(N)}$ vs. $\log{(R)}$ distribution of the clusters at the end of 23rd simulation cycle.}
\end{figure}

The morphology of the formed metal clusters has been characterized by analyzing their fractal dimension. The fractal dimension $D$ of a cluster is calculated by evaluating its center of mass and measuring the number of atoms contained inside the circumscribed sphere of radius $R$ \cite{Panshenskov2014}. The number of atoms $N$ scales with the radius $R$ as
\begin{equation}
    N(R) \sim R^D ,
\label{Eq. FractalDimension1}
\end{equation}
or equally
\begin{equation}
    \log{(N)} = D \cdot \log{(R)} + C ,
\label{Eq. FractalDimension2}
\end{equation}
where $C$ is a constant. Equation~(\ref{Eq. FractalDimension2}) enables determining the fractal dimension $D$ by considering the double-log dependence of the number of atoms in a cluster versus its circumscribed sphere radius. Results of this analysis performed for the cycle-by-cycle evolution of the most Pt-rich clusters are shown in Fig.~\ref{Fig:FractalDimension} by interconnected orange triangles. The radius of the circumscribed sphere for the largest Pt cluster stops growing after the 24th cycle when the value of $R = 6.2$~nm corresponding to the radius of the effective beam spot is reached. Green dots in Figure~\ref{Fig:FractalDimension} show the fractal dimension evaluated for all the metal clusters in the beam spot area at 23rd cycle prior to merging of separate clusters into a single structure. Blue diamonds show the corresponding values of $D$ for all the metal clusters in the beam spot area at the 30th (final) simulation cycle.
Overlapping of the values for the single cluster at different FEBID cycles and for the whole ensemble of clusters for the last cycle indicates the co-existence of clusters at various growth stages. The calculated $\log{(N)}$ values exhibit a linear dependence on $\log{(R)}$ with the fractal dimension $D = 1.55$ and the constant $C=1.76$. The evaluated value of $D$ lies in between the values $D_{{\rm 1D}} = 1$ and $D_{{\rm 2D}} = 2$ corresponding to ideal linear and planar structures, respectively, and it is within the typical range of values known for fractal-like metal clusters and aggregates \cite{Lazzari2016, Samsonov2016}.

\begin{figure}[t!]
\centering
\includegraphics[width=0.45\textwidth]{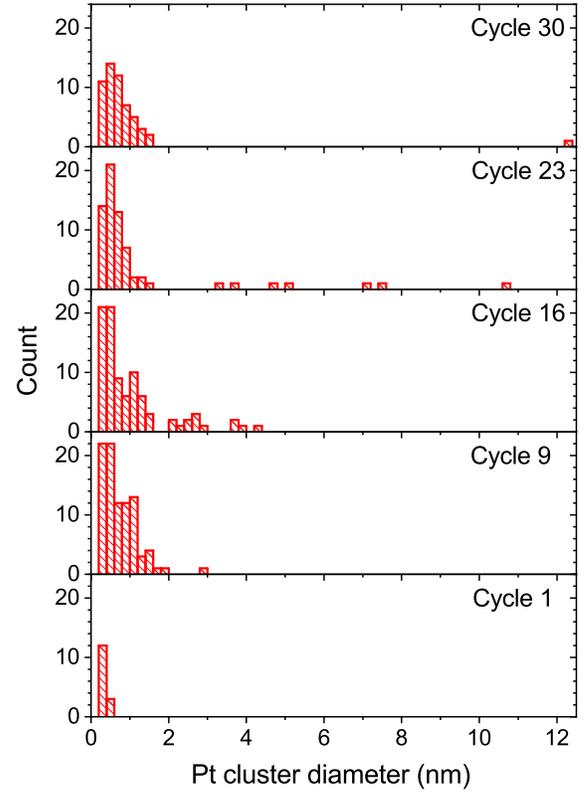}
	\caption{\label{Fig:DiameterDistribution}Distribution of the circumvented diameters $d$ of the deposited Pt clusters after several cycles of the FEBID simulation for $E_{\rm dep} = 300$~kcal/mol. }
\end{figure}

The distribution of circumvented diameters $d$ of the deposited clusters at different cycles is shown in Fig.~\ref{Fig:DiameterDistribution}. Only clusters larger than dimers 
are included in this analysis. A cluster containing $\sim$20~Pt atoms is circumscribed in a sphere of a diameter of around 1~nm. At the conditions considered in this study, separate metal clusters merge into dendritic structures. The consecutive growth of the cluster diameter is observed until the instant when most deposited atoms merge into a single nanostructure.
At this stage the nanostructure covers an effective round beam spot area with a diameter of approx. 12~nm. Further growth of the cluster will occur via an increase in the nanostructure height.

\subsection{Evaluation of nanostructure height and its metal content}

The metal content and height of nanostructures are the main experimentally measured FEBID characteristics. The experimental results \cite{Wang2004a} indicate that the growth rate and atomic content of the deposited material strongly depend on the electron fluence and the amount of adsorbed precursor molecules.

As shown in Figure~\ref{Fig:Snapshot}, the metal clusters in the simulations grow nearly isotropically in the $(xy)$-plane parallel to the substrate surface, but the cluster size and morphology depend on the radial distance from the beam center. Thus, the height and the relative Pt content of the deposited nanostructures are evaluated in concentric bins with a width of 1~nm around the beam spot axis.

\begin{figure*}[t!]
\centering
\includegraphics[width=0.92\textwidth]{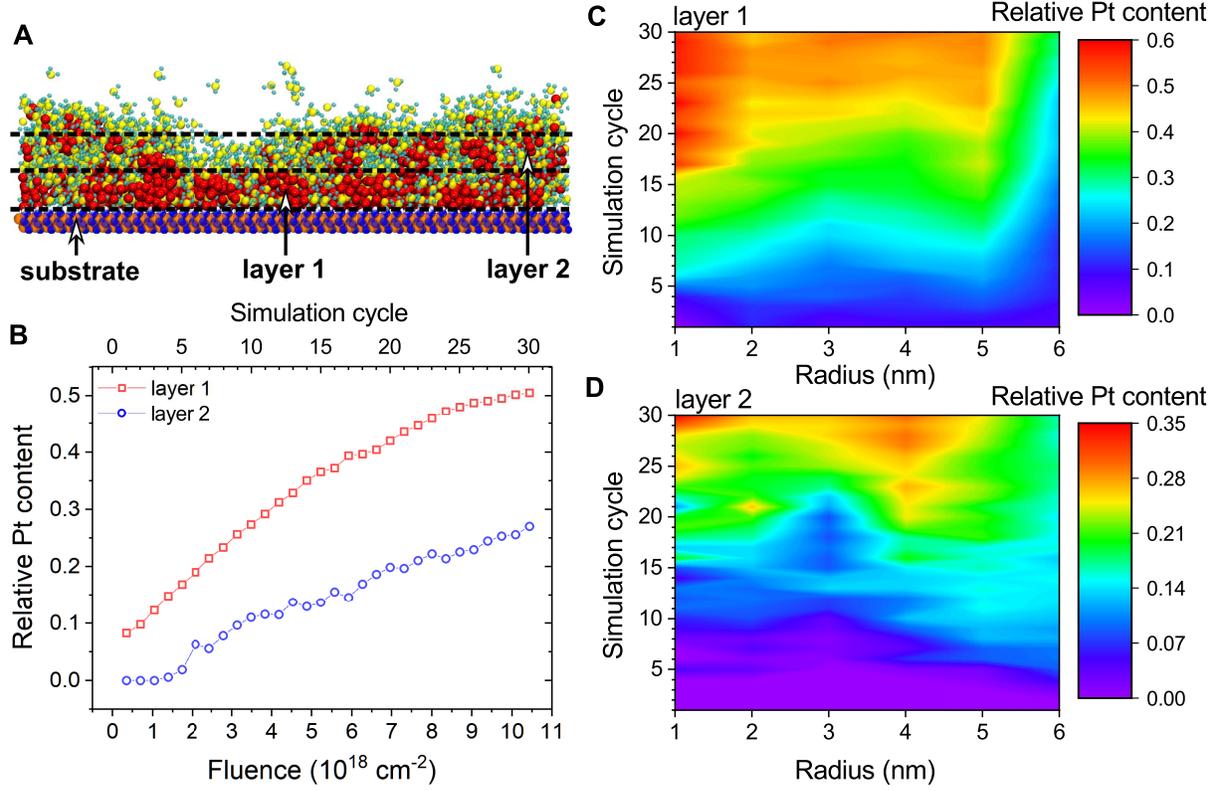}
	\caption{\label{Fig:Layer_content} (A) Atomic content of the deposited nanostructure is analyzed by splitting the nanostructure into two layers of 0.7~nm thickness each, corresponding to the height of a Pt(PF$_3$)$_4$ monolayer. (B) Relative Pt content in the beam spot area for the first two layers of 0.7~nm thickness as a function of electron fluence. Panels (C) and (D) show the evolution of the radial distribution of relative Pt content for the first and second layers, respectively. }
\end{figure*}

The nanostructure growth during the FEBID process is characterized by the atomic content of the deposited material. The relative atomic content is calculated considering all atoms in the beam spot area layer by layer. The thickness of each layer is set to 0.7~nm corresponding to the height of a Pt(PF$_3$)$_4$ monolayer adsorbed on SiO$_2$ (see Fig.~\ref{Fig:Layer_content}A). The relative Pt content is calculated by dividing the number of Pt atoms by the total number of atoms in the considered volume. The evolution of the average Pt content in the first two layers with the number of FEBID cycles is presented in Figs.~\ref{Fig:Layer_content}B--D. Panel~B shows the dependence of the relative Pt content as a function of electron fluence. Figures~\ref{Fig:Layer_content}C and D show the evolution of the radial distribution of Pt atoms for the first two layers.

Figure~\ref{Fig:Layer_content}B reveals that the relative Pt content in the first layer increases linearly during the first 11 FEBID cycles and then starts to grow slower, eventually coming to saturation.
This indicates that the formation of the first layer has been completed within the 30 FEBID cycles. Note that the first layer has undergone minor structural transformations even after the metal clusters merged into a single structure.
Further evolution of the structure will concern mainly the second and upper layers. The Pt distribution within the first layer is nearly homogeneous during the first 15 simulation cycles, while a more dense region proximal to the beam axis appears after the 15th cycle. This region corresponds to the location of the largest cluster in the simulation. Platinum atoms start filling the second layer at the 5th cycle at the edge of the beam spot.

\begin{figure*}[t!]
\centering
\includegraphics[width=0.84\textwidth]{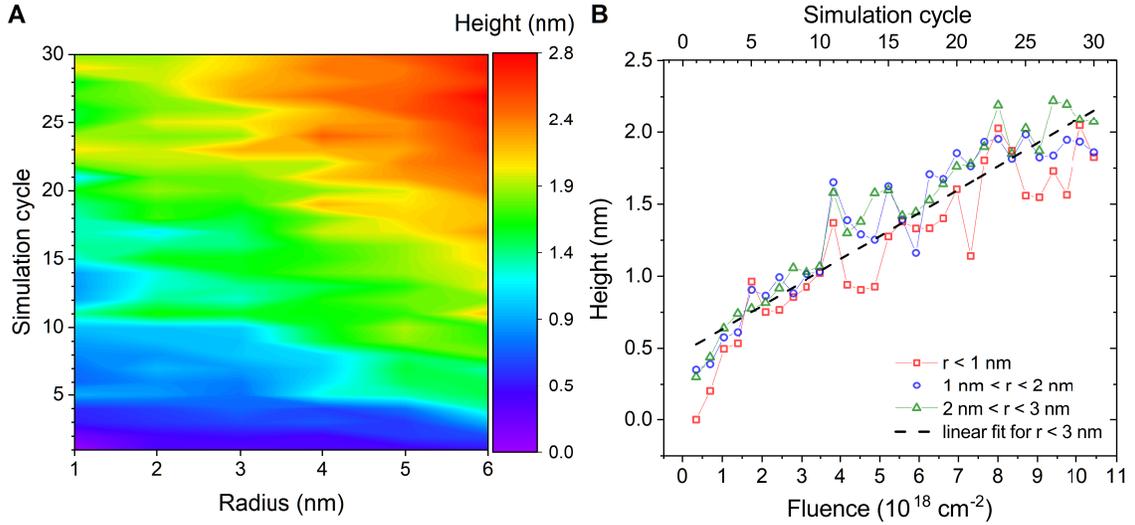}
	\caption{Evolution of the height of the grown Pt-containing structure at different FEBID simulation cycles for $E_{\rm dep} = 300$~kcal/mol. (A) Maximum height of deposited Pt atoms within concentric bins of 1~nm thickness from the electron beam axis at the end of each simulated FEBID cycle. (B) Maximum height of Pt structures within the distance of 3~nm from the electron beam axis as a function of electron fluence fitted by a linear function.}
\label{Fig:Height}
\end{figure*}

The thickness of the deposited material is calculated by the maximum $z$ coordinate of Pt atoms within the concentric bins. Figure~\ref{Fig:Height}A shows the evolution of the dependence of the nanostructure height as a function of radius from the PE beam axis in the course of the FEBID simulation. Two regions can be distinguished -- one in the center of the beam spot and another one closer to its edge (see also Fig.~\ref{Fig:Layer_content}A). The larger height of the structures at distances $4-6$~nm from the beam center arises due to the presence of attached PF$_3$ ligands, which do not allow dense packing of the Pt clusters. The height of the deposited material in the center of the beam spot (at radial distances up to 3~nm from the beam axis) as a function of electron fluence is shown in Fig.~\ref{Fig:Height}B. Considering the experimental electron flux \cite{Barry2006a} of $3.48 \times 10^{17}$~cm$^{-2}$ms$^{-1}$, the height dependence exhibits a linear behavior with the growth rate of 0.067~nm/ms. The growth rate calculated based on the data from Ref.~\cite{Barry2006a} is equal to 0.003~nm/ms. One should note that the two growth rates are evaluated at different regimes. The experimental growth rate is determined from the average linear growth of the 1758~nm high metal pillar irradiated with a stationary electron beam for 10 minutes. In the present simulations, the metal structure within the bottom-most atomic layer still undergoes minor structural transformations involving the reorganization of the metal clusters. After completing the first layer, the nanostructure growth should continue in a different regime with the growth rate being comparable to the experimental value. The difference between the experimental value of the growth rate and the value derived from the simulations can also be ascribed to different rates of the precursor deposition and the larger value of $E_{\rm dep}$.

\section{Conclusions}
\label{Conclusions}

This study has presented a general protocol for the systematic atomistic modeling of the FEBID using the MBN Explorer and MBN Studio software packages. The protocol is based on the methodology developed previously \cite{Sushko2016, DeVera2020} and is applicable to any combination of precursor, substrate and electron beam parameters.
Using the protocol, the irradiation-driven molecular dynamics (IDMD) simulations of the nanostructure growth can be performed for various irradiation and replenishment regimes corresponding to realistic experimental conditions.

The proposed computational methodology is applicable to characteristic system sizes and time scales that can be treated by the classical molecular dynamics approach. The model requires the specification of the following input data: interatomic potentials describing the interaction of atoms in the system, precursor fragmentation cross sections, and distributions of primary, secondary and backscattered electrons. The outcome of the FEBID process is governed by a balance of different processes, such as adsorption, desorption, diffusion and fragmentation of the precursor molecules. The cumulative contribution of all these processes defines the regime at which the FEBID goes. The presented methodology reproduces the molecular system's state before, during and after irradiation, without explicit simulations of the all aforementioned processes. The model permits the inclusion of other phenomena such as chemisorption or electron stimulated desorption. The presented method is under continuous development, so a more detailed and accurate description of the essential processes will be included in future studies. Finally, while being focused on the FEBID process using a pulsed electron beam, the methodology can be adjusted to simulate the nanostructure formation by other nanofabrication techniques using electron beams, such as direct-write electron beam lithography.

Atomistic simulations provide the complete characterization of the morphology and internal structure of metal deposits. An early stage of the FEBID process (nucleation, growth and coalescence of the metal clusters) involves the initial formation of the deposited layer, which drives the further growth of the material.
Atomistic analysis of the simulation results provides space-resolved relative metal content, height and the growth rate of the deposits, and the linear size of the clusters, which represent valuable reference data for the experimental characterization of the FEBID material.

The presented simulation workflow has been successfully utilized to analyze the nanostructure formation with Pt(PF$_3$)$_4$ precursor molecules. The performed simulation of 30 FEBID cycles corresponds to an early stage of the nanostructure formation and the creation of several metal-enriched layers. At this stage small metal-containing clusters start to grow, merge and interconnect, forming branched structures.
The process of metal atoms nucleation and the formation of a single metal structure can be correlated with time-dependent electrical conductivity measurements during FEBID \cite{Porrati2009, Hochleitner2008}. Coalescence of smaller metal clusters into a single structure should correspond to a jump in the electrical current through the deposit.
The calculated fractal dimension of the metal structures is equal to 1.55, which is a typical value for the nanostructures created by metal cluster deposition. The metal structure formed after merging separate clusters covers the effective beam spot area with a diameter of 12 nm.
The average Pt content in the beam spot area is approximately 30\%. These findings can be verified experimentally with the STM analysis.


\section*{Acknowledgements}
This work was supported by the Deutsche Forschungsgemeinschaft (Project no. 415716638), and the European Union's Horizon 2020 research and innovation programme -- the Radio-NP project (GA 794733) within the H2020-MSCA-IF-2017 call and the RADON project (GA 872494) within the H2020-MSCA-RISE-2019 call. This work was also supported in part by the COST Action CA17126 ``Towards understanding and modelling intense electronic excitation'' (TUMIEE). The authors gratefully acknowledge the possibility to perform computer simulations at Goethe-HLR cluster of the Frankfurt Center for Scientific Computing. The authors are also grateful to Prof. Michael Huth for fruitful discussions.

\bibliography{MBN-RC}

\end{document}